\documentclass[prl,twocolumn,superscriptaddress,showpacs,unsortedaddress]{revtex4}
\usepackage{graphicx}
\usepackage{dcolumn}
\usepackage{amsmath}
\usepackage{hyperref}
\usepackage{bm}
\usepackage{float}
\usepackage{tabularx}
\usepackage{color}

\begin{document}

\title{A Photon Regeneration Experiment for Axionlike Particle Search using X-rays}

\author{R. Battesti}
\email{remy.battesti@lncmi.cnrs.fr}
\author{M. Fouch\'e}
\affiliation{Laboratoire National des Champs Magn\'etiques Intenses (UPR 3228, CNRS-INSA-UJF-UPS), F-31400 Toulouse Cedex, France, EU}

\author{C. Detlefs}
\author{T. Roth}
\affiliation{European Synchrotron Radiation Facility, F-38043 Grenoble, France, EU}

\author{P. Berceau}
\author{F. Duc}
\author{P. Frings}
\author{G.L.J.A. Rikken}
\author{C. Rizzo}
\affiliation{Laboratoire National des Champs Magn\'etiques Intenses (UPR 3228, CNRS-INSA-UJF-UPS), F-31400 Toulouse Cedex, France, EU}

\date{\today}

\begin{abstract}
In this letter we describe our novel photon regeneration experiment
for the axionlike particle search using a x-ray beam with a photon
energy of $50.2\,\mathrm{keV}$ and $90.7\,\mathrm{keV}$, two
superconducting magnets of 3\,T, and a Ge detector with a high
quantum efficiency. A counting rate of regenerated photons
compatible with zero has been measured. The corresponding limits on
the pseudoscalar axionlike particle-two photon coupling constant is
obtained as a function of the particle mass. Our setup widens the
energy window of purely terrestrial experiments devoted to the
axionlike particle search by coupling to two photons. It also opens
a new domain of experimental investigation of photon propagation in
magnetic fields.
\end{abstract}


\maketitle


Photon propagation in magnetic fields is a long standing domain of
research for QED tests \cite{Iacopini} and for particle searches
beyond the standard model \cite{Maiani1986}. All the experiments
performed up to now have used a photon energy of the order of 1\,eV
(see \cite{BMV} and Refs. therein). Higher photon energies have been
proposed to increase the signal, in particular $\gamma$\,rays
\cite{CantatoreGamma} for QED test, or to increase the parameter
space for particle searches, in particular x-rays
\cite{Rabadan,Dias}.

As far as particle searches are concerned, photon regeneration
experiments \cite{Sikivie1,Sikivie2,VanBibber}, also called ``light
shining through the wall" experiments, are an important tool in the
search for massive particles that couple to photons in the presence
of magnetic fields. Such particles are predicted by many extensions
of the standard model. A very well known example is the standard
axion, a pseudoscalar chargeless boson proposed to solve the strong
CP problem \cite{PecceiQuinn,Weinberg,Wilczek} i.e.~the difference
between the value of the neutron electric dipole moment predicted by
QCD and its experimental value \cite{Baker}.

The principle of a photon regeneration experiment is to send a
polarized photon beam through a region where a transverse magnetic
field is present, and then to stop the photons by a wall. Since they
hardly interact with matter, axionlike particles (ALPs) generated in
the magnetic region upstream of the wall can pass through it. Behind
the wall, a second magnetic field region allows to convert back ALPs
into photons. Several photon regeneration experiments have been
performed \cite{BRFT1,BRFT2,BMV2,GammeV,LIPSS,OSCQAR,ALPS2010}: none
of them has ever detected regenerated photons. They have therefore
set limits on the ALP-two photon coupling constant $g$ and the
particle mass $m_\mathrm{a}$. The best limits can be found in
Ref.~\cite{ALPS2010}.

Limits are usually given for masses $m_\mathrm{a} \ll \omega$
\cite{Units}, where $\omega$ is the photon energy, but a detailed
theoretical analysis of ALP-photon and photon-ALP conversion
amplitudes valid for $m_\mathrm{a} \leq\omega$ can be found in
Ref.\,\cite{Adler}. Again, for all the photon regeneration
experiments performed up to now, $\omega$ is of the order of
$1\,\mathrm{eV}$. Experiments searching for ALPs of astrophysical
origin, such as ADMX \cite{ADMX} and CAST \cite{CAST}, provide
better limits than the purely terrestrial ones. ADMX looks for
galactic cold dark matter $\mu $eV ALP conversion into microwave
photons in a resonant cavity immersed in a static magnetic field,
while CAST looks for axions or ALPs generated in
the core of the sun. These ALPs travel to earth and are converted
back into photons of a few keV in a static laboratory magnetic
field. Due to the higher photon energy, the CAST limits extend up to
masses on the order of a few eV \cite{CAST}. These limits, however,
depend on the model used to calculate the flux of ALPs to be
detected. The critical sensitivity to these models is exposed by the
recent proposal of an ALP with a 17\,meV mass which could explain
the observed spectral shape of the x-ray solar emission
\cite{Zioutas}. In this case ALPs coming from the sun's interior
would be reconverted into photons near the sun's surface, thus
escaping the detection by CAST.

Increasing the photon energy in photon regeneration experiments
allows to test new regions of the $m_\mathrm{a}$ and $g$ parameter
space.  The use of soft x-rays has been proposed in
Ref.\,\cite{Rabadan}, namely at the VUV-FEL free electron laser at
DESY, providing photons of energy between $10\,\mathrm{eV}$ and
$200\,\mathrm{eV}$. The use of hard x-rays from a synchrotron light
source has been proposed in Ref.~\cite{Dias}. Synchrotron light
sources provide photons with energy of several tens of keV, much
higher than the photon energy available nowadays at free electron
lasers.

\begin{figure}[h]
\begin{center}
\resizebox{1\columnwidth}{!}{
\includegraphics{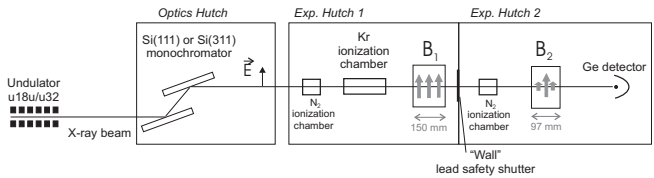}}
\caption{\label{Fig:Setup} Experimental Setup. The double crystal
monochromator is adjusted to select the desired photon energy. The
first experimental hutch corresponds to the ALP generation area with
the transverse magnetic field $B_1$. The second experimental hutch
contains the second magnetic field $B_2$ which allows to reconvert
ALPs to photons. These photons are detected by a liquid nitrogen
cooled Ge detector with a high quantum efficiency. Ionization
chambers placed along the beam path measure the incident flux or
serve for alignement purposes. The synchrotron x-rays are polarized
parallel to the magnetic fields.}
\end{center}
\end{figure}

In this letter we describe our photon regeneration experiment using
x-ray beams with a photon energy of $50.2\,\mathrm{keV}$ and
$90.7\,\mathrm{keV}$, carried out at the European Synchrotron
Radiation Facility (ESRF), France, on beamline ID06
\cite{Monochromatic}. Our setup consists of two superconducting
magnets that provide magnetic fields of 3 T over a length of 150 and
97\,mm respectively, and a Ge detector with a high quantum
efficiency for the stated photon energies. This configuration widens
the energy domain probed by purely terrestrial ALP searches. A
counting rate of regenerated photons compatible with zero has been
measured. We present the corresponding limits on the ALP-two photon
coupling constant as a function of the particle mass. Thanks to the
high photon energy, our limits extend to a parameter region where no
model independent limits have been set so far. In particular our
experimental results provide limits on the existence of
$17\,\mathrm{meV}$ ALPs.


Our experimental setup is shown in Fig.\,\ref{Fig:Setup}. We use two
different photon energies, $\omega = 50.2\,\mathrm{keV}$ and
$90.7\,\mathrm{keV}$, corresponding to slightly different settings
of the x-ray beamline. For $50.2\,\mathrm{keV}$ (resp.
$90.7\,\mathrm{keV}$), a Si(111) (resp. Si(311)) double crystal
monochromator is adjusted to select x-rays emitted by the
$5^{\mathrm{th}}$ (resp. $9^\mathrm{th}$) harmonic of the cryogenic
permanent magnet multipole undulator source U18, closed to a gap of
$6.0\,\mathrm{mm}$ \cite{PhD_Kitegi,Chavanne2010}. The energy bandwidth is 7.3\,eV (resp.
6.8\,eV). For both energies, the size of the beam is $2 \times
2\,\mathrm{mm^2}$ and the synchrotron x-rays are horizontally
polarized. The beam direction is stabilized by a feedback loop
adjusting the pitch of the second monochromator crystal to ensure a
position stability better than 0.1\,mm at the entrance of the second
magnet.

Most of the beam path is under vacuum in order to avoid air
absorption. The incident flux is measured thanks to ionization
chambers filled with 1\,bar of nitrogen or krypton. Different
ionization chambers placed along the beam path let us check for any
photon loss due to beam misalignment for example. During data
acquisition, the 30\,cm long krypton filled ionization chamber,
located just before the first magnet, is used to precisely monitor
the incident flux. The beamline has
delivered about $1.2\times 10^{12}$ photons per second at 50.2\,keV
and $3.1\times 10^{10}$ photons per second at 90.7\,keV.

The magnetic fields are provided by two superconducting magnets with
the field direction parallel to the x-ray polarization, the
experiment being thus sensitive to pseudoscalar particles
\cite{Scalar}. Their diameter aperture is about 2\,cm and the
pressure inside the magnets is less than $10^{-4}$\,mbar. Both
magnets have been manufactured by Oxford Instruments. The first one
has provided a maximum magnetic field $B_1 = 3$\,T which can be
regarded as uniform along the beam path over a length of $L_1 =
150$\,mm. The second magnet was lent to us by the DUBBLE beamline
(BM26) \cite{Bras2003} at the ESRF. It has also delivered $B_2 =
3$\,T. The shape of its magnetic field along the beam direction can
be approximated by a triangular shape with a half base length of
$L_2 = 97$\,mm, the maximum of 3\,T being at the center of the
magnet.

The magnets are located separately in the two lead shielded
experimental hutches, EH1 and EH2 respectively, of the beamline. The
safety shutter between EH1 and EH2 serves as the wall to block the
x-ray beam. It consists of a 50\,mm thick lead plate. Similarly, the
x-ray regeneration and detection section is shielded by the
radiation hutch EH2. The complete enclosure of the primary x-ray
beam in EH1 and the additional shielding of EH2 lead to a
comfortably low level of x-ray background radiation dominated by
cosmic events.

The detection system is based on a 5\,mm thick Ge detector (Canberra
GL0055) cooled with liquid nitrogen. The sensitive area is 6\,mm in
diameter. X-ray photons arriving on the detector create electric
charges proportional to the photon energy, which are amplified
(Canberra 2024) and filtered by a single channel analyzer (Ortec
850) to reject events that do not correspond to the photon energy
selected by the monochromator. This detection system combines an
acceptable quantum efficiency of $\approx 99.98\%$ at
$50.2\,\mathrm{keV}$ and $\approx 84\%$ at $90.7\,\mathrm{keV}$,
with a reasonably low dark count rate. This background count rate
was measured at $(7.2\pm1.4)\times 10^{-3}$\,photons per second
while the x-ray beam was turned off, as shown on the first line of
Table\,\ref{Table:Results}. The error corresponds to 95\,$\%$
confidence level.


\begin{table*}
  \centering
  \begin{tabular*}{1\textwidth}{@{\extracolsep{\fill}} c c c c c c c}
  \hline
   X-ray beam & Magnets & $\omega$ (keV) & $t_{\mathrm{i}}$ (s) & $N_{\mathrm{inc}}$ (Hz) & $N_{\mathrm{c}}$ (Hz) & $N_\mathrm{p}$ (Hz)\\
  \hline
  \hline
  OFF & OFF & & 13913 & 0 & $(7.2\pm 1.4)$ $\times 10^{-3}$ &\\
  ON & OFF & 50.2 & 7575 & $1.2\times 10^{12}$ & $(5.7 \pm 1.8)$ $\times 10^{-3}$ &\\
  ON & ON & 50.2 & 7276 & $1.2\times 10^{12}$ & $(6.2\pm 1.8)$ $\times 10^{-3}$ & $(0.5\pm 2.6)\times 10^{-3}$\\
  ON & OFF & 90.7 & 7444 & $3.2\times 10^{10}$ & $(7.9 \pm 2.0)$ $\times 10^{-3}$ &\\
  ON & ON & 90.7 & 7247 & $3.1\times 10^{10}$ & $(8.1 \pm 2.2)$ $\times 10^{-3}$ & $(0.2\pm 3.0)\times 10^{-3}$\\
  \hline
\end{tabular*}
  \caption{
  Summary of our data acquisition taken with magnets on or off,
  x-ray beam on or off. The integration time is denoted as $t_\mathrm{i}$,
  while $N_\mathrm{inc}$ is the number of incident photons per
  second, $N_{\mathrm{c}}$ is the number of detected photons
  per second and $N_\mathrm{p}$ is number of regenerated photons per second.
  Errors correspond to 95\,$\%$ confidence level. No excess count rate above
  background has been detected.}\label{Table:Results}
\end{table*}

The following experimental protocol is used before each data
acquisition. First, the monochromator is adjusted to select the
desired energy while keeping an incident flux as high as possible.
Then, the detector is moved about 20\,cm sideways from the direct
beam position. The safety shutter is opened, allowing the x-ray beam
to propagate through both experimental hutches. In this position,
the dominant radiation received at the detector are photons
elastically scattered by air \cite{Compton}. This is used to adjust
the upper and lower thresholds of the single channel analyzer such
that only photons of the selected energy are counted. The upper
(lower) threshold is 10\,$\%$ above (20\,$\%$ below) the voltage
generated by the elastically scattered photons. Next, the detector
is protected by Cu absorbers and it is moved back into the direct
beam position to check its geometrical alignment. Finally, before
data collection the safety shutter is closed and the Cu absorbers
are removed. The procedure is repeated after data collection.


Results are summarized in Table\,\ref{Table:Results}. The
integration time $t_\mathrm{i}$ is about 2 hours for each photon
energy in two different configurations -- with or without the
magnetic fields. The count rate $N_\mathrm{c}$ is the number of
photons detected per second. The error on $N_\mathrm{c}$ corresponds
to 95\,$\%$ confidence level and is given by
$2\sqrt{N_\mathrm{c}/t_\mathrm{i}}$ since the distribution of the
detected photons is a Poisson distribution. The number of
regenerated photons per second $N_\mathrm{p}$ is the difference
between count rates measured with and without the magnetic fields.
We see that no excess count above the background level has been
detected. Finally the upper photoregeneration probability at
95\,$\%$ confidence level corresponds to the error on $N_\mathrm{p}$
over the incident photon rate $N_\mathrm{inc}$. It is $P =
2.2\times10^{-15}$ at 50.2\,keV, and $P = 9.7\times10^{-14}$ at 90.7
keV.

The photon to ALP conversion and reconversion transition probability
after propagating in vacuum over a distance $z$ in an inhomogeneous
magnetic field $B$ may be written as \cite{Raffelt1988}:
\begin{equation}
p\left(z\right) = \left| \int_0^z dz' \Delta_g\left(z'\right) \times
\exp(i \Delta_\mathrm{a} z') \right|^2, \label{eq:P_integral}
\end{equation}
where $\Delta_g(z) = \frac{gB(z)}{2} \quad \mbox{and}\quad
\Delta_\mathrm{a} = - \frac{m_\mathrm{a}^2}{2\omega}$. Finally, the
photoregeneration probability is:
\begin{equation}
P = \eta p_1p_2,
\label{eq:PhotoRegProb}
\end{equation}
with $\eta$ the detection efficiency, $p_1$ the conversion
probability in the first magnet and $p_2$ the reconversion
probability in the second magnet. These equations are correct for
$m_\mathrm{a} \ll \omega$.

\begin{figure}[h]
\begin{center}
\resizebox{1\columnwidth}{!}{
\includegraphics{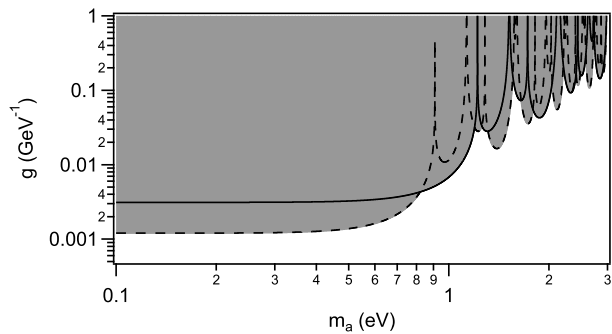}}
\caption{\label{Fig:XAX_results} 95\,$\%$ confidence level limits on
the ALP-two photon coupling constant $g$ as a function of the
particle mass $m_{\mathrm{a}}$. The grey area is excluded. The
dashed line represents limits obtained with a photon energy of
$50.2$\,keV while the solid line corresponds to $90.7$\,keV.}
\end{center}
\end{figure}

Our experimental sensitivity limit for the ALP-two photon coupling
constant versus mass is calculated by numerically solving
Eqs.\,(\ref{eq:P_integral}) and (\ref{eq:PhotoRegProb}), using the
upper photon regeneration probability experimentally measured. To
this end, the real profiles of the magnetic fields along the beam
direction provided by the manufacturers are used. Our limits at
95\,$\%$ confidence level are plotted in
Fig.\,\ref{Fig:XAX_results}. In particular, $g<1.3\times
10^{-3}$\,GeV$^{-1}$ for masses lower than 0.4\,eV, and $g<6.8\times
10^{-3}$\,GeV$^{-1}$ for masses lower than 1\,eV. Our limits could
be extended up to 90\,keV \cite{Adler}, but because of the phase
mismatching they decrease very rapidly when
$m_\mathrm{a}\gg\sqrt{\omega/L_{1,2}}$, thus becoming less
interesting. Moreover, for such masses the probability oscillates so
rapidly that its actual value depends critically on the exact value
of the experimental parameters $L_{1,2}$ and $\omega$. In this case
the level of confidence of corresponding limits is mostly limited by
the confidence level on these experimental values. We believe that a
detailed discussion of this issue is out of the scope of our letter.

\begin{figure}
\begin{center}
\resizebox{1\columnwidth}{!}{
\includegraphics{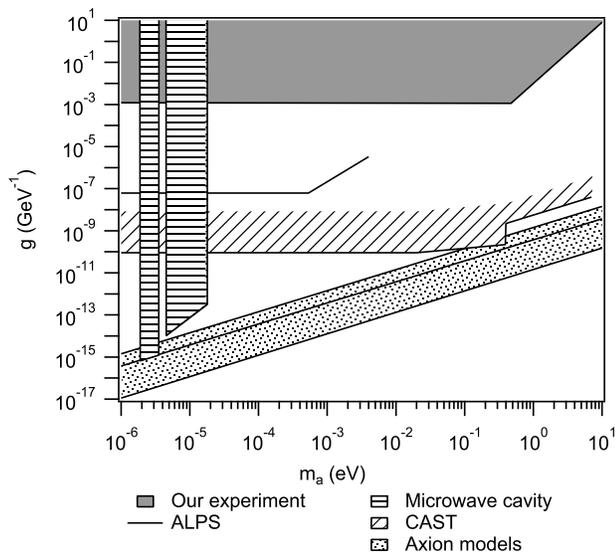}
} \caption{Limits on the ALP-two photon coupling constant $g$ as a
function of the particle mass $m_\mathrm{a}$ obtained by
experimental searches. Our exclusion region is presented as the grey
area. See text for more details.} \label{fig:CourbeGenerale_Axion}
\end{center}
\end{figure}

We compare our limits to other limits obtained with laboratory
experiments in Fig.\ref{fig:CourbeGenerale_Axion}. Our exclusion
region is presented as the grey area. The best limits obtained on a
purely laboratory experiment by the ALPS collaboration
\cite{ALPS2010} with a 95\,$\%$ confidence level is the region above
the solid line. The best limits set by the search of
extraterrestrial ALPs are the two hashed areas, namely the 95\,$\%$
confidence level exclusion region of CAST (diagonally hashed)
\cite{CAST}, and the 90\,$\%$ confidence level exclusion region on
microwave cavity experiments (horizontally hashed)
\cite{CavityRBF1,CavityRBF2,CavityUF,ADMX}. Model predictions
\cite{AxionModel} are also shown as a dotted stripe (line in
between: $E/N=0$ \cite{KSVZ1,KSVZ2}). This figure shows that we have
tested a new region of the $m_\mathrm{a}$ and $g$ parameter space
for purely terrestrial -- model independent -- experiments.

Our experiment could certainly be upgraded. A longer acquisition
time would improve the limits, but an improvement of a factor of 2
requires a 16 times longer acquisition. This also applies for the
photon flux and for detector noise rate. The latter could likely be
improved by using the x-ray detector in anticoincidence with cosmic
ray detectors put around it and/or in coincidence with the electron
bunches circulating in the synchrotron ring. Using higher magnetic
fields increases limits linearly, which is obviously more
interesting. A static 15 T field can be reasonably envisaged. Longer
magnets could provide higher limits but only at low masses since
longer magnets reduce the coherence length of the photon-ALP
oscillations and limits at higher masses. The best solution would be
to increase the magnetic field $B$ and reduce the magnet length $L$
keeping the product $B \times L$ as high as possible.

Our experiment extends the search of photon oscillations into
massive particles in the presence of magnetic fields to higher
energies. The observed low background count rate clearly
demonstrates the sensitivity of ``shining through the wall"
experiments with a synchrotron light source. Moreover we studied for
the first time the propagation of x-ray photons in magnetic fields
opening a new domain of experimental investigations.

\acknowledgments This work has been performed in the framework of
the BMV project. We acknowledge the ESRF for providing beam time on
ID06 and financial support. The detection system has been kindly
provided by the ESRF Detector Pool. We thank J.-P.\,Nicolin for his
technical support. We thank W. Bras for kindly lending us one of the
two superconducting magnets, P. van der Linden for its technical
support and C.\,Cohen and M.\,Kocsis for their help with the Ge
detector. Finally we thank A. Dupays and J. Jaeckel for fruitful
discussions. We gratefully acknowledge the support of the
\textit{Fondation pour la recherche IXCORE}.

\bibliography{XAX}

\begin{thebibliography}{38}
\bibitem[{Uni()}]{Units}
\bibinfo{note}{Natural Lorentz-Heaviside units with $\hbar = c = 1$ are
  employed throughout.}
\bibitem[{Mon()}]{Monochromatic}
\bibinfo{note}{A white x-ray beam could also be used. But our experiment was
  performed on an existing station that can operate with monochromatic beam
  only.}
\bibitem[{Sca()}]{Scalar}
\bibinfo{note}{To be sensitive to scalar particles, the direction of the first
  magnetic field has to be turned in order to be perpendicular to the X-ray
  polarization.}
\bibitem[{Com()}]{Compton}
\bibinfo{note}{Compton scattering, which would change the energies of the
  scattered photons arriving on the detector, can be neglected at the energy
  and the angles used in this experiment.}
\end{thebibliography}

\end{document}